\documentclass[sigconf, anonymous=False]{acmart}
\usepackage{tabularx, booktabs}
\usepackage{graphicx}
\usepackage{multirow}
\usepackage{subcaption}
\usepackage{array}
\usepackage{balance}
\usepackage{enumitem}
\usepackage{balance}
\usepackage{diagbox}
\usepackage{dsfont}

\newcolumntype{C}[1]{>{\centering\arraybackslash}p{#1}}
\newcolumntype{R}[1]{>{\raggedleft\arraybackslash}p{#1}}
\newcolumntype{L}[1]{>{\raggedright\arraybackslash}p{#1}}

\makeatletter
\newcommand{\ssymbol}[1]{^{\@fnsymbol{#1}}}
\makeatother

\makeatletter
\def\hlinew#1{%
  \noalign{\ifnum0=`}\fi\hrule \@height #1 \futurelet
   \reserved@a\@xhline}
\makeatother

\AtBeginDocument{%
  \providecommand\BibTeX{{%
    \normalfont B\kern-0.5em{\scshape i\kern-0.25em b}\kern-0.8em\TeX}}}

\setcopyright{acmcopyright}
\copyrightyear{2021}
\acmYear{2021}
\acmDOI{}
\acmISBN{}

\acmConference[SIGIR'21]{The 44th International ACM SIGIR Conference on Research and Development in Information Retrieval}{July 11-15, 2021}{}

\begin{document}
\title{Optimizing Dense Retrieval Model Training with Hard Negatives}


\author{Jingtao Zhan}
\affiliation{BNRist, DCST, Tsinghua University}
\email{jingtaozhan@gmail.com}

\author{Jiaxin Mao}
\affiliation{GSAI, Renmin University of China}
\email{maojiaxin@gmail.com}

\author{Yiqun Liu}
\authornote{Corresponding author}
\affiliation{BNRist, DCST, Tsinghua University}
\email{yiqunliu@tsinghua.edu.cn}

\author{Jiafeng Guo}
\affiliation{University of Chinese Academy of Sciences, Institute of Computing Technology, CAS}
\email{guojiafeng@ict.ac.cn}

\author{Min Zhang}
\affiliation{BNRist, DCST, Tsinghua University}
\email{z-m@tsinghua.edu.cn}

\author{Shaoping Ma}
\affiliation{BNRist, DCST, Tsinghua University}
\email{msp@tsinghua.edu.cn}

\renewcommand{\shortauthors}{Zhan, et al.}


\begin{abstract}
Ranking has always been one of the top concerns in information retrieval researches. For decades, the lexical matching signal has dominated the ad-hoc retrieval process, but solely using this signal in retrieval may cause the vocabulary mismatch problem. In recent years, with the development of representation learning techniques, many researchers turn to Dense Retrieval (DR) models for better ranking performance. Although several existing DR models have already obtained promising results, their performance improvement heavily relies on the sampling of training examples. Many effective sampling strategies are not efficient enough for practical usage, and for most of them, there still lacks theoretical analysis in how and why performance improvement happens. To shed light on these research questions, we theoretically investigate different training strategies for DR models and try to explain why hard negative sampling performs better than random sampling. Through the analysis, we also find that there are many potential risks in static hard negative sampling, which is employed by many existing training methods. Therefore, we propose two training strategies named a Stable Training Algorithm for dense Retrieval (STAR) and a query-side training Algorithm for Directly Optimizing Ranking pErformance (ADORE), respectively. STAR improves the stability of DR training process by introducing random negatives. ADORE replaces the widely-adopted static hard negative sampling method with a dynamic one to directly optimize the ranking performance. Experimental results on two publicly available retrieval benchmark datasets show that either strategy gains significant improvements over existing competitive baselines and a combination of them leads to the best performance.
\end{abstract}



\maketitle

\section{introduction}

Document ranking is essential for many IR related tasks, such as question answering~\cite{guu2020realm} and Web search~\cite{craswell2020overview, bajaj2016ms}.  
An effective ranking algorithm will benefit many downstream tasks related to information access researches~\cite{han2020learning, nogueira2019document}.  
Traditional algorithms such as BM25~\cite{robertson1994some} usually utilize exact term matching signals. Their capabilities are therefore limited to keyword matching and may fail if the query and document use different terms for the same meaning, which is known as the vocabulary mismatch problem. 
To better deduce the users' search intent and retrieve relevant items, the ranking algorithms are expected to conduct semantic matching between queries and documents~\cite{li2014semantic}, which is a challenging problem. 

In recent years, with the development of deep learning~\cite{vaswani2017attention, devlin2019bert, liu2019roberta}, especially representation learning techniques~\cite{bengio2013representation}, many researchers have turned to the Dense Retrieval (DR) model to solve the semantic matching problem~\cite{luan2020sparsedense, karpukhin2020dense, lin2020distilling, guu2020realm}. 
In essence, DR attempts to encode queries and documents into low-dimension embeddings to better abstract their semantic meanings.  
With the learned embeddings, document index can be constructed and the query embedding can be adopted to perform efficient similarity search for online ranking.
Previous studies showed that DR models achieve promising results on many IR-related tasks~\cite{karpukhin2020dense, ding2020rocketqa, guu2020realm}.

However, there are some unsolved but essential problems related to DR's effectiveness and training efficiency\footnote{We focus on the training efficiency because the efficiency in the inference process is guaranteed by the maximum inner product search algorithms\cite{shen2015learning, johnson2019billion}.}. 
Firstly, though previous works achieved promising results using different training strategies, their empirical conclusions are sometimes different and even conflict with each other. 
For example, researchers have conflicting opinions on whether training with hard negatives outperforms that with random negatives~\cite{huang2020embedding, xiong2020approximate, ding2020rocketqa, luan2020sparsedense}. 
Secondly, many existing well-performed training methods are relatively inefficient and not so applicable for practical usage. 
For example, some previous works utilize computationally expensive methods, such as knowledge distillation~\cite{lin2020distilling} and periodic index refresh~\cite{guu2020realm, xiong2020approximate} in the training process. 

We believe that these problems are mainly caused by the lack of theoretical understanding of the training process for DR models. A fair comparison among the existing training strategies will help us design more effective and efficient optimization methods. Therefore, we theoretically formulate the DR training methods and investigate the relationship between training methods and the target optimization objectives. 

Through the analysis, we find that random negative sampling aims to minimize the total pairwise errors.
Thus, it has a critical problem that some difficult queries can easily dominate its training process and result in serious top-ranking performance loss.
On the contrary, hard negative sampling minimizes the top-K pairwise errors. Therefore, it is more suitable for improving the ranking performance of top results, which is also the target of many popular IR systems, such as Web search engines.

Furthermore, we look into two kinds of hard negatives, namely static and dynamic ones.
The static ones are employed by many existing training methods. These methods adopt a traditional retriever~\cite{gao2020complementing, karpukhin2020dense, luan2020sparsedense} or a warm-up DR model~\cite{xiong2020approximate, guu2020realm} to pre-retrieve the top results as unchanging hard negatives during training.
Theoretical analysis shows that their loss functions cannot guarantee performance improvement, and experiments show their training process is unstable.
The dynamic ones are the real hard negatives and rely on the current model parameters. During training, they are dynamically changing because the model is updated iteratively.
We show that they resolve the problems of static ones and thus are better for training DR models.

To improve the existing training strategies and help DR models gain better performance, we propose two methods to effectively train the DR models.
Firstly, we propose a \textbf{S}table \textbf{T}raining \textbf{A}lgorithm for dense \textbf{R}etrieval (STAR) to stabilize the static hard negative sampling method. 
It utilizes static hard negatives to optimize the top-ranking performance and random negatives to stabilize the training. It reuses the document embeddings in the same batch to improve efficiency. Experimental results show that STAR outperforms competitive baselines with significant efficiency improvement.
Secondly, we propose a method to train DR models with dynamic hard negative samples instead of static ones, named a query-side training \textbf{A}lgorithm for \textbf{D}irectly \textbf{O}ptimizing \textbf{R}anking p\textbf{E}rformance (ADORE). 
It adopts the document encoder trained by other methods and further trains the query encoder to directly optimize IR metrics.
Therefore, it can be used to improve other methods by training a better query encoder. 
Experimental results show that ADORE significantly boosts the ranking performance. With the help of ADORE, some simple training methods can even match the existing competitive approaches in terms of effectiveness and greatly outperform them in terms of efficiency.   
The combination of ADORE and STAR achieves the best performance and is still much more efficient than current popular training methods~\cite{xiong2020approximate, guu2020realm}.
We further investigate using index compression techniques~\cite{jegou2010product} to greatly save computational resources for ADORE. 

In summary, our contributions are as follows:
\begin{itemize}
	\item We theoretically investigate the training process of DR models and compare different popular training strategies, including random sampling and hard negative sampling. Our analysis reveals that these strategies lead to different optimization targets and that hard negative sampling better optimizes the top-ranking performance. Experimental results verify our theoretical analysis results.
	\item We investigate one of the most popular training strategies for DR models that employ static hard negative samples. We theoretically and empirically demonstrate its potential risks in decreasing ranking performance. 
	\item We propose two training strategies that employ hard negatives to optimize DR models. Experimental results on two popular retrieval benchmark datasets show that they achieve significant performance improvements. Their combination achieves the best retrieval performance\footnote{Code, trained models, and retrieval results are available at \url{https://github.com/jingtaozhan/DRhard}.}.
\end{itemize}


\section{Related Work}
\label{sec:related_work}


Dense Retrieval represents queries and documents with embeddings.
During the offline stage, it encodes documents and builds the document index. During the online stage, it encodes the input queries and performs similarity search~\cite{johnson2019billion}. 
Researchers mainly use negative sampling methods to train DR models except the recently proposed knowledge distillation method~\cite{lin2020distilling}. We introduce them in the following. 

Several works utilized random negative sampling for training DR models. \citet{huang2020embedding} used random negative samples to approximate the recall task. \citet{karpukhin2020dense} and \citet{zhan2020repbert} adopted In-Batch training to use other queries' relevant documents in the same mini-batch as negatives, which we believe is approximately random negative sampling. 
\citet{ding2020rocketqa} found it is beneficial to increase the number of random negatives in the mini-batch.

Some works applied hard-negative mining to train DR models. \citet{gao2020complementing} and \citet{karpukhin2020dense} used BM25 top documents as hard negatives. \citet{xiong2020approximate} and \citet{guu2020realm} used a warm-up DR model to pre-retrieve the top documents as hard negatives during training. They also periodically re-build the index and refresh the hard negatives, which greatly increases the computational cost. Since the hard negatives are fixed during the entire training process or for a few thousand training steps, we refer to these methods as static hard negative sampling.

However, some researchers found static hard negative sampling does not help or even harms the ranking performance. \citet{luan2020sparsedense} found it brings no benefits for the open-domain QA task. \citet{ding2020rocketqa} even found it significantly worsens the ranking performance. 

Although \citet{xiong2020approximate} tried to theoretically analyze the training process of DR models based on convergence speed, we find their conclusions contradict the experimental results and will discuss it in Section~\ref{sec:exp_random_vs_hard}.
On the contrary, we present a more fundamental analysis based on optimization objectives, and the conclusions are well supported by the experiments.

\section{Task Formulation}
\label{sec:task_formulation}

Given a query $q$ and a corpus $\mathcal{C}$, the DR model with parameters $\theta$ is to find the relevant documents $D^+$. 
It encodes queries and documents separately into embeddings, denoted as $\vv{X}_{q;\theta}$ and $\vv{X}_{d;\theta}$, respectively. 
Then it uses the similarity function, often inner product, to perform efficient retrieval.
Let $f(q, d)$ be the predicted relevance score. It equals to:
\begin{equation}
f(q, d) = \langle \vv{X}_{q;\theta} , \vv{X}_{d;\theta} \rangle
\end{equation}
where $\langle,\rangle$ denotes the similarity function. 

DR models are typically trained with pairwise loss where each training sample consists of a query, a negative document $d^-$, and a positive document $d^+$~\cite{zhan2020repbert, ding2020rocketqa, xiong2020approximate}. 
For ease of explanation, we use the following pairwise loss function:
\begin{equation}
\mathcal{L}(d^+,d^-) = \mathds{1}_{f(q, d^+) < f(q, d^-)}
\label{eq:forward}
\end{equation}
where $\mathds{1}_{A}$ is an indicator function, which is $1$ if $A$ holds and 0 otherwise. Therefore, we can establish the relationship between the ranking position of the positive document $\pi(d^+)$ and the training loss with respect to it as follows:
\begin{equation}
\pi(d^+) = \delta(d^+) + 1 + \sum_{d^- \in D^-} \mathcal{L}(d^+,d^-)
\label{eq:Rank}
\end{equation}
where $\delta(d^+)$ is the number of relevant documents ranked higher than $d^+$ and $D^-$ is all the irrelevant documents, i.e., $\mathcal{C} \backslash D^+$.

In practice, we cannot directly optimize over all the samples in a corpus since the cost is prohibitive. Therefore, an important question is what distribution should the negative documents be sampled from. 
Given a query $q$, different sampling strategies can be viewed as setting different weights $w(d^-)$ for each negative document $d^-$. 
Therefore, the general form of the learning objective is as follows: 
\begin{equation}
\theta^* = \arg\min_{\theta} \sum_{q} \sum_{d^+ \in D^+} \sum_{d^- \in D^-} w(d^-) \cdot \mathcal{L}(d^+,d^-)  
\label{eq:unified_training}
\end{equation}

\section{Random vs. Hard Negatives}
\label{sec:benefit_hard_neg}
This section provides a theoretical explanation of how hard negatives help optimize retrieval models. 

\subsection{Random Negative Sampling}
We start with introducing the widely-used random negative sampling method~\cite{huang2020embedding, xiong2020approximate, ding2020rocketqa}. 
It uniformly samples negatives from the entire corpus.
Using the formulation in Section~\ref{sec:task_formulation}, we can see that using random negatives is equivalent to minimizing $\pi(d^+)$ (or minimizing the total pairwise errors):
\begin{equation}
\begin{aligned}
{\theta}^* &= \arg\min_{\theta} \sum_{q} \sum_{d^+ \in D^+} \sum_{d^- \in D^-} \mathcal{L}(d^+,d^-) \\
		 &= \arg\min_{\theta} \sum_{q} \sum_{d^+ \in D^+} \pi(d^+) - \delta(d^+) - 1 \\
		 &= \arg\min_{\theta} \sum_{q} {\rm const} + \sum_{d^+ \in D^+} \pi(d^+) \\
\end{aligned}
\label{eq:random_negative_sampling}
\end{equation}

While minimizing the total pairwise errors seems reasonable, we argue that the above optimization objective has the following critical problem. 
The loss function with respect to a single query $q$ is \emph{unbounded}. $\pi(d^+)$ could be as large as the size of the whole corpus $|\mathcal{C}|$. Therefore, the overall loss will be dominated by the queries with large $\pi(d^+)$, and the model can hardly focus on improving top-ranking performance. 
Section~\ref{sec:exp_random_vs_hard} will show that this problem leads to serious performance loss in practice.

Therefore, random negative sampling is sub-optimal for training DR models and it is necessary to investigate how to optimize the retrieval performance of DR models with other sampling strategies.

\subsection{Hard Negative Sampling}
\label{sec:subsec_hard_neg}
Another sampling strategy is to sample top-K documents as negatives. 
This paper refers to them as hard negatives.
The optimization process is formulated as follows.
\begin{equation}
\begin{aligned}
{\theta}^* &= \arg\min_{\theta} \sum_{q} \sum_{d^+ \in D^+} \sum_{d^- \in D^-} \mathds{1}_{\pi(d^-) \leq K_q} \cdot \mathcal{L}(d^+,d^-) \\
		 &= \arg\min_{\theta} \sum_{q} \sum_{d^+ \in D^+} \min(\pi(d^+)-\delta(d^+)-1, K) \\
\end{aligned}
\label{eq:hard_negative_sampling}
\end{equation}
where $K$ is the number of hard negative documents,
and $K_q$ is the ranking position of the top $K_{th}$ negative document in $D^-$, i.e., $\sum_{d^- \in D^-} \mathds{1}_{\pi(d^-) \leq K_q} = K$.

Comparing Eq.~(\ref{eq:hard_negative_sampling}) with Eq.~(\ref{eq:random_negative_sampling}), we can see that hard negative sampling minimizes the top-K pairwise errors instead of the total pairwise errors and the loss w.r.t. a single query is bounded by $K$.
In this sense, the adoption of hard negatives alleviates the unbounded loss problem of random negative sampling and leads to a more robust optimization in retrieval performance. 
Hard negative sampling emphasizes the top-ranking performance and disregards the lower-ranked pairs that hardly affect the user experience or evaluation metrics. Therefore, it is more in line with the truncated evaluation metrics. 

The following theorem shows the relationship between Eq.~(\ref{eq:random_negative_sampling}) and (\ref{eq:hard_negative_sampling}) in terms of whether and when two sampling strategies lead to the same optimal parameter weights.  
\begin{theorem}
\label{theorem:hard_and_rand}
Let ${\theta}_h^*$ be the optimal parameters for Eq.~(\ref{eq:hard_negative_sampling}). If  $\forall d^+, \pi(d^+)-\delta(d^+)-1 \leq K$, then ${\theta}_h^*$ is also the optimal parameters for Eq.~(\ref{eq:random_negative_sampling}).
\end{theorem}
That is to say, if DR models can rank all relevant documents at high positions, using random negative sampling and hard negative sampling in training DR models will result in the same optimal parameters.  
However, DR models may not be effective for some training queries, especially those that require keyword match~\cite{zhan2020repbert, gao2020complementing, lin2020distilling}. In this case, the random negative sampling strategy would pay too much attention to the difficult queries, which, however, would not be reflected by the popular adopted truncated evaluation metrics.

\section{Static vs. Dynamic Hard Negatives}
\label{sec:risk_static_hardneg}
In Section~\ref{sec:benefit_hard_neg}, we have shown that hard negative sampling is more effective than random negative sampling. 
Since the real hard negatives are dynamically changing during training, we refer to them as dynamic hard negatives. 
However, the current main strategy is using a traditional retriever~\cite{gao2020complementing, karpukhin2020dense, luan2020sparsedense} or a warm-up DR model~\cite{xiong2020approximate, guu2020realm} to pre-retrieve the top documents as fixed hard negatives during training. Thus, we refer to them as static hard negatives\footnote{
Some works~\cite{xiong2020approximate, guu2020realm} periodically refresh the index and retrieve static hard negatives during training.
Though we do not consider it in the theoretical analysis, we will explore it in our experiments and show its limited capability to resolve the problems analyzed in this section. 
}.
Next, we will theoretically discuss these two kinds of hard negatives.

\subsection{Static Hard Negatives}
\label{sec:subsec_static_hard_neg}
We formulate static hard negative sampling as follows. 
\begin{equation}
\begin{aligned}
{\theta}^* &= \arg\min_{\theta} \sum_{q} \sum_{d^+ \in D^+} \sum_{d^- \in D^-} \mathds{1}_{d^- \in D_s^-} \cdot \mathcal{L}(d^+,d^-) 
\end{aligned}
\label{eq:static_hard_negative_sampling}
\end{equation}
where $D_s^-$ is a set of pre-retrieved hard negatives $(|D_s^-| \ll |\mathcal{C}|)$. 
To characterize $D_s^-$, we define its `quality' $\phi(D_s^-)$ as the highest ranking position:
\begin{equation}
\begin{aligned}
\phi(D_s^-)  = \min_{d^- \in D_s^-}\pi(d^-) \in [1, |\mathcal{C}|-|D_s^-|+1] \\
\end{aligned}
\label{eq:static_hard_neg_unstable}
\end{equation}

We can see that $\phi(D_s^-)$ is very loosely bounded and theoretically can be very large. 
Though previous works~\cite{gao2020complementing, karpukhin2020dense, luan2020sparsedense, xiong2020approximate, guu2020realm} implicitly assumed that $\phi(D_s^-)$ is always very small during the entire training process, we argue that is not necessarily true.
The training process is likely to drive $\phi(D_s^-)$ large instead of small because the gradient-based optimization constantly forces the DR model to predict small relevance scores for $D_s^-$. Therefore, $D_s^-$ is likely to be ranked lower and lower during training, which, however, is invisible to the loss. Section~\ref{sec:exp_static_dynamic} will verify that the DR model quickly ranks $D_s^-$ very low.

Using the above notions, we discuss the ranking performance of an `ideal' DR model whose training loss is minimum zero.
We use mean reciprocal rank (MRR) to evaluate the ranking performance. The infimum of MRR is as follows: 
\begin{equation}
\begin{aligned}
& \sum_{q} \sum_{d^+ \in D^+} \sum_{d^- \in D^-} \mathds{1}_{d^- \in D_s^-}  \cdot \mathcal{L}(d^+,d^-) = 0  \\
\Rightarrow  {\rm inf}({\rm MRR}) &= E_q \frac{1}{\phi(D_s^-) - |D^+|} \geq \frac{1}{|\mathcal{C}|-|D_s^-|+1-  |D^+|} \approx \frac{1}{|\mathcal{C}|} \\
\end{aligned}
\label{eq:static_hard_neg_unstable}
\end{equation}
where $\rm inf$ denotes the infimum. 

Since $\phi(D_s^-)$ can be very large as discussed above, the infimum can be almost zero and thus the top-ranking performance is not theoretically guaranteed.
In the worst case, MRR is approximately $\frac{1}{|\mathcal{C}|}$.
Considering that we are discussing an `ideal' DR model whose training loss is zero, this result is unacceptable.
Therefore, we believe it is risky to use static hard negative sampling.


\subsection{Dynamic Hard Negatives}
\label{sec:subsec_benefit_against_static_hard_neg}
This section will show the benefits of dynamic hard negatives compared with static ones. The dynamic hard negatives are actually the top-ranked irrelevant documents given the DR parameters at any training step.
For ease of comparison, we formally define the dynamic hard negatives $D^-_\theta$ based on Eq.~(\ref{eq:hard_negative_sampling}):
\begin{equation}
\begin{aligned}
D^-_\theta = \{ d^-: \mathds{1}_{\pi(d^-) \leq K_q} = 1 \}
\end{aligned}
\end{equation}
During training, $D^-_\theta$ is dynamically changing because $\theta$ is constantly updated. 
In fact, the theoretical analysis in Section~\ref{sec:subsec_hard_neg} is based on dynamic hard negatives and we have shown their benefits compared with random negatives. 

Since we use $\phi(D_s^-)$ to represent the `quality' of static hard negatives, we show the `quality' of $D^-_\theta$ for comparison.  Considering that $D^-_\theta$ always contains the top-ranked negatives, it is well bounded:
\begin{equation}
\begin{aligned}
\phi(D^-_\theta) = \min_{d^- \in D^-_\theta}\pi(d^-) \in [1, |D^+| + 1] 
\end{aligned}
\end{equation}
Therefore, according to Eq.~(\ref{eq:static_hard_neg_unstable}), MRR achieves the maximum one if the training loss achieves the minimum zero:
\begin{equation}
\begin{aligned}
 & \sum_{q} \sum_{d^+ \in D^+} \sum_{d^- \in D^-} \mathds{1}_{\pi(d^-) \leq K_q} \cdot \mathcal{L}(d^+,d^-) = 0 \\
\Rightarrow \quad & \phi(D^-_\theta) = |D^+| + 1 \quad \Rightarrow {\rm \quad MRR} = 1
\end{aligned}
\end{equation}

In this sense, dynamic hard negative sampling is better than the static one.

\section{Training Algorithms}
\label{sec:training_algorithms}
In Section~\ref{sec:benefit_hard_neg} and \ref{sec:risk_static_hardneg}, we show that random negative sampling and static hard negative sampling have problems in terms of effectiveness and training stability, respectively. 
Thus, we propose two improved training strategies for better training DR models, namely STAR and ADORE. 


\begin{figure}
    \hspace*{\fill}%
    \subcaptionbox{Input: one relevant doc and multiple static hard negatives are sampled for each query.
    \label{fig:star_input}}
    {\includegraphics[height=.28\linewidth]{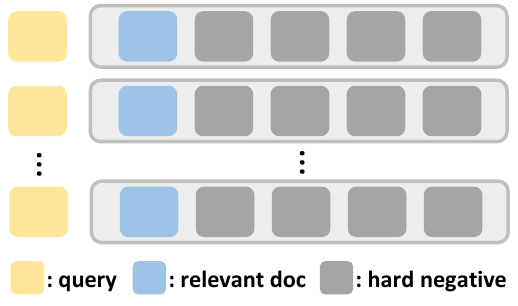}}
    \hfill\hfill\hfill\hfill%
    \subcaptionbox{Reusing other document embeddings when computing pairwise loss.
    \label{fig:star_loss}}
    {\includegraphics[height=.28\linewidth]{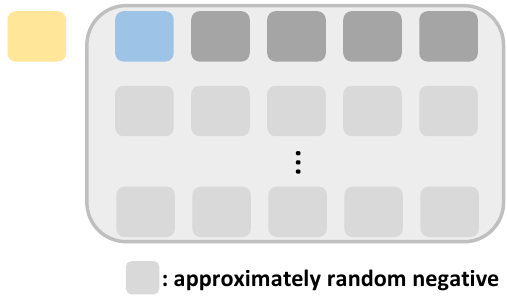}}%
    \hspace*{\fill}%
    \caption{Reusing strategy of STAR. The number of rows correspond to the batch size.}
\end{figure}

\subsection{STAR}
According to our analysis in Section~\ref{sec:risk_static_hardneg}, static hard negative sampling is unstable. Therefore, we propose a \textbf{S}table \textbf{T}raining \textbf{A}lgorithm for dense \textbf{R}etrieval (STAR). STAR aims to employ static hard negatives to improve top-ranking performance and random negatives to stabilize the training process. Moreover, it reuses the document embeddings in the same batch to greatly improve efficiency.


\subsubsection{Employing Static Hard Negatives}
STAR uses a warm-up model to retrieve the top documents for all training queries, which serve as the static hard negatives and will not be updated. We use them to approximate the real dynamic hard negatives. 
The inspiration comes from our pilot experiment where we find different well-trained DR models tend to recall the same set of documents but with different ranking order. 

\subsubsection{Employing Random Negatives}
To stabilize the training process, STAR additionally introduces random negatives during training and optimizes the following objective:
\begin{equation}
{\theta}^* = \alpha \cdot L_r(\theta) + L_s(\theta) \quad (0\!<\!\alpha\!\ll\!1)
\end{equation}
where $L_r(\theta)$ is the loss of random negative sampling defined as Eq.~(\ref{eq:random_negative_sampling}), $L_s(\theta)$ is the loss of static hard negative sampling defined as Eq.~(\ref{eq:static_hard_negative_sampling}), and $\alpha$ is a hyper-parameter.
If static hard negatives well approximate the dynamic ones, $L_s(\theta)$ tends to be large and dominates the training process.
If not, $L_s(\theta)$ tends to be small and STAR approximately degenerates into a random negative sampling method, which is much better than the worst case of static hard negative sampling. 

\subsubsection{Improving Efficiency}
STAR adopts a reusing strategy to improve efficiency. 
It does not explicitly add random negative documents to the input and instead reuse other documents in the same batch as approximately random negatives. 
Figure~\ref{fig:star_input} shows one input batch of STAR, which is the same as static hard negative sampling method~\cite{xiong2020approximate}. Each row has one query, its relevant document, and its static hard negative documents. 
Figure~\ref{fig:star_loss} shows how STAR computes pairwise loss for one query. The document embeddings in other rows are also utilized as negatives, which we believe can approximate the random negatives. 
The reusing strategy could also be regarded as an improved version of In-Batch training strategy~\cite{zhan2020repbert, karpukhin2020dense}.

\subsubsection{Loss Function}
STAR adopts RankNet pairwise loss~\cite{burges2010ranknet}. Given a query $q$, let $d^+$ and $d^-$ be a relevant document and a negative document. $f(q,d)$ is the relevance score predicted by the DR model. The loss function $\mathcal{L}_R$ is formulated as follows:
\begin{equation}
\mathcal{L}(d^+,d^-) = {\rm log}(1+e^{f(q, d^-)-f(q, d^+)})
\label{eq:star_loss}
\end{equation}

\begin{figure}
    \hspace*{\fill}%
    \subcaptionbox{Negative sampling method. 
    \label{fig:negative_flowchart}}
    {\includegraphics[height=.6\linewidth]{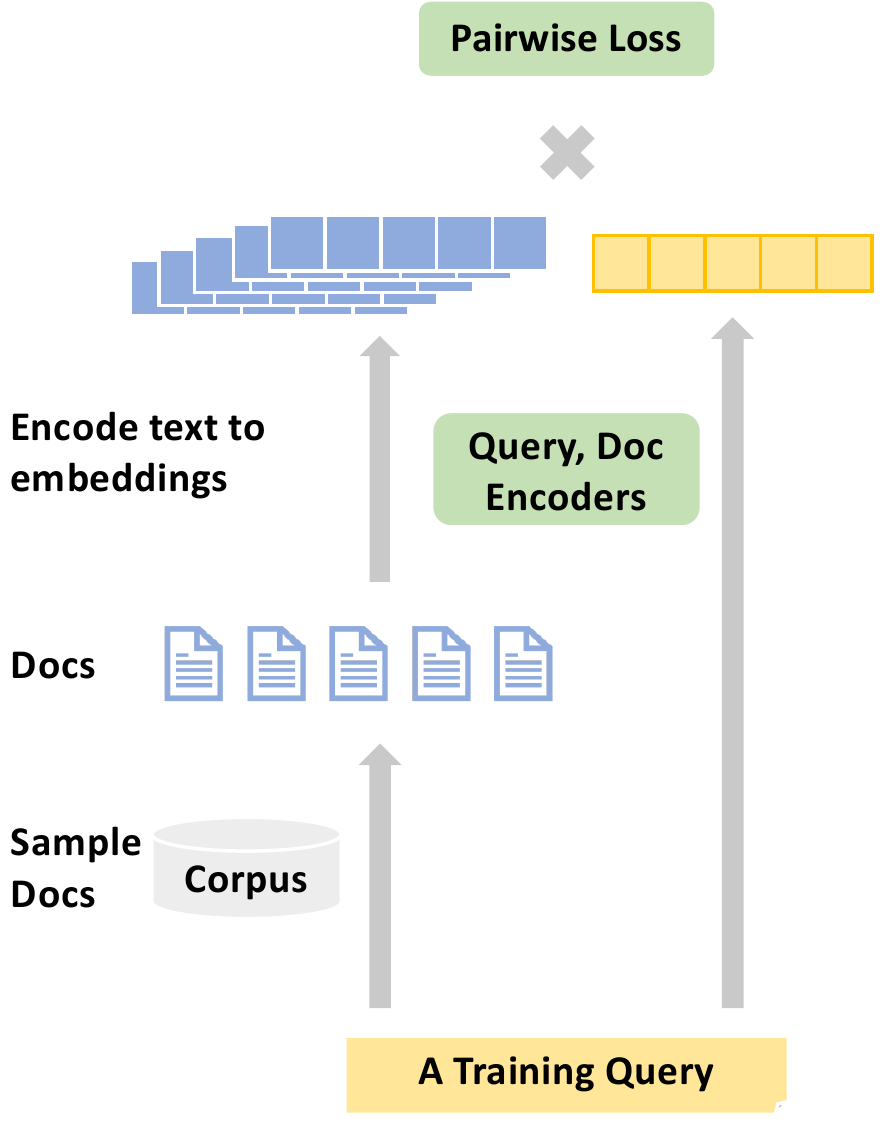}}
    \hfill%
    \subcaptionbox{ADORE.
    \label{fig:adore_flowchart}}
    {\includegraphics[height=.6\linewidth]{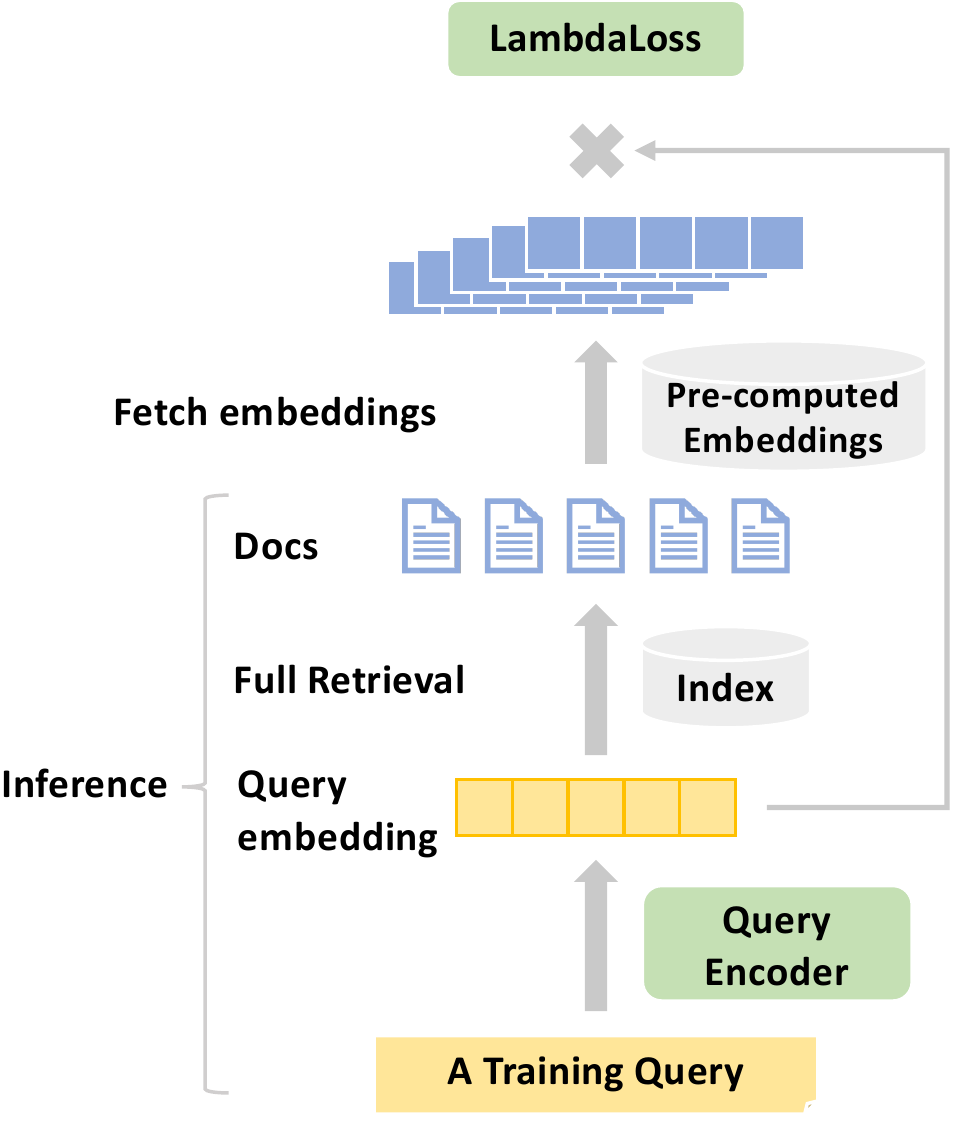}}%
    \hspace*{\fill}%
    \caption{The flow chart of negative sampling training method and our proposed ADORE. Batch size is set to one.}
\end{figure}
\subsection{ADORE}
Our previous analysis in Section~\ref{sec:benefit_hard_neg} and \ref{sec:risk_static_hardneg} shows great potential of utilizing dynamic hard negatives for training DR models.
Therefore, this section presents a query-side training \textbf{A}lgorithm for \textbf{D}irectly \textbf{O}ptimizing \textbf{R}anking p\textbf{E}rformance (ADORE).
It utilizes dynamic hard negatives and LambdaLoss~\cite{burges2010ranknet} to directly optimize ranking performance.  
It adopts a pre-trained document encoder and trains the query encoder.
We show the training process of ADORE in Figure~\ref{fig:adore_flowchart} and the common negative sampling method in Figure~\ref{fig:negative_flowchart} for comparison. 

\subsubsection{Employing Dynamic Hard Negatives} 
ADORE resolves the difficulty to acquire dynamic hard negatives as follows. Before training, ADORE pre-computes the document embeddings with a pre-trained document encoder and builds the document index. They are fixed throughout the entire training process. 
At each training iteration, it encodes a batch of queries and uses the embeddings to retrieve the corresponding top documents, which are the real dynamic hard negatives. 
ADORE utilizes them to train the DR model.
To the best of our knowledge, ADORE is the first DR training method to employ dynamic hard negatives. Its effectiveness is guaranteed by the theoretical analysis in Section~\ref{sec:subsec_hard_neg} and \ref{sec:subsec_benefit_against_static_hard_neg}. 

\subsubsection{Directly Optimizing IR Metrics}
Unlike previous methods, ADORE performs retrieval at each step and thus can utilize the listwise approach~\cite{liu2009learning} to directly optimize the ranking performance.  
We use LambdaLoss~\cite{burges2010ranknet} to derive a weight $w(d^-)$ for each training pair to better optimize IR metrics. 
Given a query $q$, a positive document $d^+$, and a negative document $d^-$, let $\Delta \mathcal{M}$ be the size of the change in target IR metric given by swapping the ranking positions of $d^+$ and $d^-$. Through multiplying $\Delta \mathcal{M}$ to RankNet loss, \citet{burges2010ranknet} empirically showed that it could directly optimize IR metrics. Therefore, ADORE employs the following loss functions:
\begin{equation}
\label{eq:lambdaloss}
\mathcal{L}(d^+,d^-) = \Delta \mathcal{M} \cdot {\rm log}(1+e^{f(q, d^-)-f(q, d^+)})  
\end{equation}
According to \citet{qin2010general}, when the training set is large enough, this method will yield the optimal model in terms of the corresponding metric. 

\subsubsection{End-to-end Training} 
\label{sec:adore_end_to_end}
Document index is often compressed in practice to save computational resources. 
Different compression techniques may lead to different optimal DR parameters and hence should be considered during training.
Though previous methods ignore this information, ADORE can well perform end-to-end training with the actual document index used in inference as shown in Figure~\ref{fig:adore_flowchart}.
In this regard, ADORE alleviates the discrepancy between training and inference and can achieve better ranking performance. We will empirically verify this in section~\ref{sec:exp_effectiveness_adore}.

\subsection{Combining STAR and ADORE}
Both ADORE and STAR have their own advantages.
ADORE directly optimizes the ranking performance while STAR cannot. 
STAR optimizes both the query encoder and the document encoder while ADORE only optimizes the query encoder.
We combine the two strategies by using STAR to train the document encoder and using ADORE to further train the query encoder.

\section{Experimental Settings}
\label{sec:exp_setting}

\subsection{Datasets}

We conduct experiments on the TREC 2019 Deep Learning~(DL) Track~\cite{craswell2020overview}. The Track focuses on ad-hoc retrieval and consists of the passage retrieval and document retrieval tasks. 
The passage retrieval task has a corpus of 8.8 million passages with 503 thousand training queries, 7 thousand development queries, and 43 test queries. 
The document retrieval task has a corpus of 3.2 million documents with 367 thousand training queries, 5 thousand development queries, and 43 test queries. 
We use the official metrics to evaluate the top-ranking performance, such as MRR@10 and NDCG@10. Besides, R@100 is adopted to evaluate the recall power.

\subsection{Baselines}

\subsubsection{Sparse Retrieval} 
We list several representative results according to the TREC overview paper~\cite{craswell2020overview} and runs on MS MARCO~\cite{bajaj2016ms} leaderboard, such as BM25~\cite{yang2018anserini}, the best traditional retrieval method, BERT weighted BM25~(DeepCT)~\cite{dai2019context}.

\subsubsection{Dense Retrieval} 
\label{sec:DR_baseline} 
The DR baselines include several popular training methods. 
For random negative sampling baselines, we present Rand Neg~\cite{huang2020embedding} and In-Batch Neg~\cite{karpukhin2020dense, zhan2020repbert}. The former randomly samples negatives from the entire corpus, and the latter uses other queries' relevant documents in the same batch as negative documents.
For static hard negative sampling baselines, we present BM25 Neg~\cite{gao2020complementing} and ANCE~\cite{xiong2020approximate}.
BM25 Neg uses the BM25 top candidates as the negative documents. 
ANCE uses a warm-up model to retrieve static hard negatives. Every $10k$ training steps, it encodes the entire corpus, rebuilds the document index, and refreshes the static hard negatives with the current DR model parameters. 
For knowledge distillation baseline, we present TCT-ColBERT~\cite{lin2020distilling} which uses ColBERT~\cite{Khattab2020ColBERTEA} as the teacher model.

\subsubsection{Cascade IR}
Although this paper focuses on the retrievers, we employ cascade systems for further comparison. We report the performances of the best LeToR model and the BERT model~\cite{nogueira2019passage}, which use BM25 as the first-stage retriever.

\subsection{Implementation Details}
All DR models use the $\text{RoBERTa}_\text{base}$~\cite{liu2019roberta} model as the encoder. 
The output embedding of the ``[CLS]'' token is used as the representation of the input text. 
We use the inner product to compute the relevance score and adopt the Faiss library~\cite{johnson2019billion} to perform the efficient similarity search.
Documents are truncated to a maximum of 120 tokens and 512 tokens for the passage and document tasks, respectively. 
The top-200 documents are used as the hard negatives. 

The implementation details for DR baselines are as follows.
In-Batch Neg and Rand Neg models are trained on passage task with Lamb optimizer~\cite{you2019large}, batch size of $256$, and learning rate of $2 \times 10^{-4}$. 
We find LambdaLoss~\cite{burges2010ranknet} cannot bring additional performance gains and hence use the RankNet loss.
The trained models are directly evaluated on the document task because this produces better retrieval performance~\cite{yan2019idst, xiong2020approximate}. 
We use the open-sourced BM25 Neg model and ANCE model~\cite{xiong2020approximate}. They are re-evaluated so we can perform the significance test. Minor performance variances are observed compared with the originally reported values.
Since TCT-ColBERT~\cite{lin2020distilling} is not open-sourced, we borrow the results from its original paper and do not perform the significance test on it.

STAR uses the BM25 Neg model as the warm-up model, which is the same as ANCE and hence their results are directly comparable.
It uses Lamb optimizer, batch size of $256$, and learning rate of $1 \times 10^{-4}$ on passage task.
It uses AdamW optimizer~\cite{loshchilov2017decoupled}, batch size of $60$, and learning rate of $2\times 10^{-6}$ on document task.

ADORE uses AdamW optimizer, learning rate of $5 \times 10^{-6}$, and batch size of $32$ on both tasks.
$\mathcal{M}$ in Eq.~(\ref{eq:lambdaloss}) is MRR@200 and MRR@10 on passage and document tasks, respectively.  
ADORE can improve a trained DR model by further training its query encoder. 
For example, ADORE+Rand Neg means it uses the document encoder trained by Rand Neg and further trains the query encoder.

\section{Experimental Results}
\label{sec:experiment_results}
We conduct experiments to verify our theoretical analysis and the effectiveness of our proposed methods. Specifically, this section studies the following research questions:
\begin{itemize}
	\item \textbf{RQ1:} How do random negatives and hard negatives affect optimization objectives?
	\item \textbf{RQ2:} How does the static hard negative sampling method perform in practice? 
	\item \textbf{RQ3:} How effective and efficient are our proposed training methods?
\end{itemize}

\begin{figure}
    \includegraphics[width=0.7\linewidth, keepaspectratio=True]{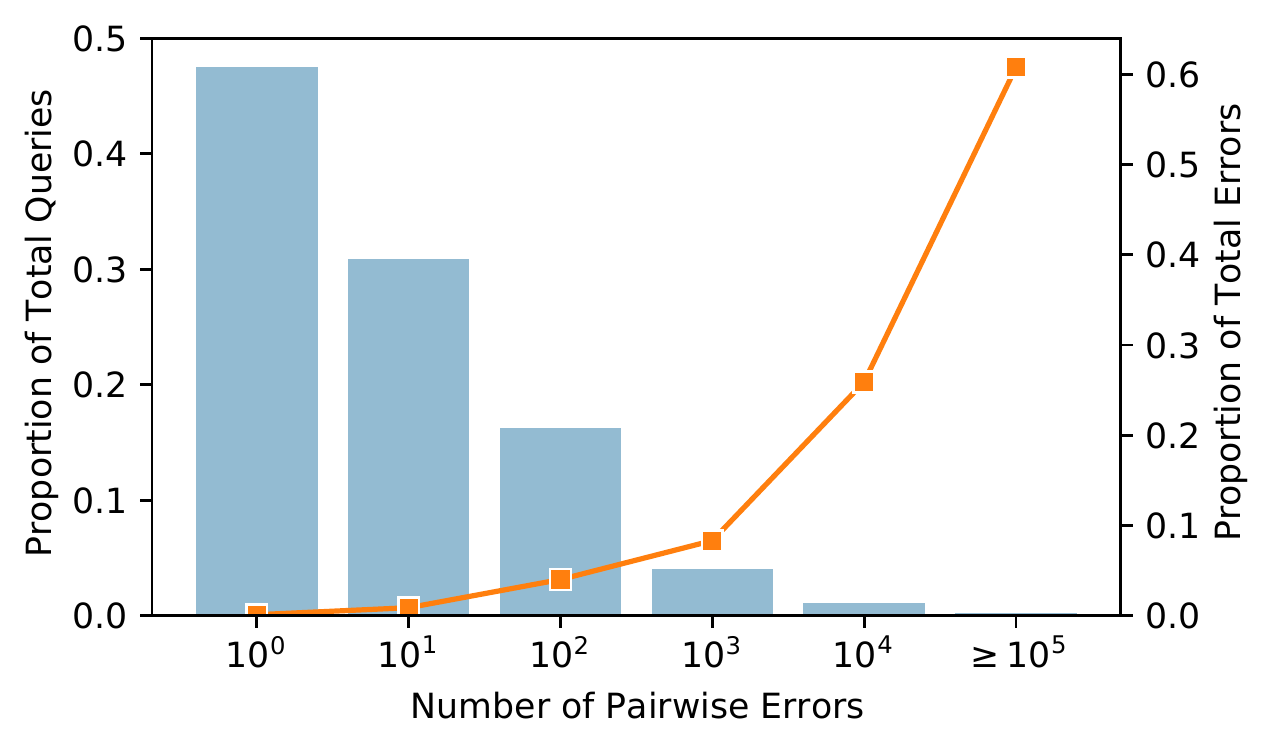}
    \caption{Distribution of pairwise errors per query on MARCO Dev Passage dataset for Rand Neg model. The histogram shows the proportion of total queries, and the line chart shows the proportion of total pairwise errors. 
    } 
    \label{fig:dominate}
\end{figure}

\begin{table}[t]
    \centering
    \caption{Total pairwise errors and top-K pairwise errors of different DR models. Best results are marked bold. K equals to 200.}
    ~\label{tab:optimize_target}
    \begin{tabular}{l|cc} \hlinew{0.8pt}
    \textbf{Models} & \textbf{Total Errors} & \textbf{Top-K Errors} 
    \\ \hline
    
In-Batch Neg	& \phantom{0}679  & 43.2 \\
Rand Neg 	& \textbf{\phantom{0}659}  & 39.3 \\
BM25 Neg 	& 2432 & 46.4 \\
ANCE 		& 1448 & 37.3 \\
STAR 	& 1128 & \textbf{35.8} \\
\hline
ADORE+In-Batch Neg 	& \textbf{\phantom{0}649}  & 37.7 \\
ADORE+Rand Neg 	& \phantom{0}736  & 36.8 \\
ADORE+BM25 Neg 	& 1840 & 40.0 \\
ADORE+ANCE 		& 1345 & 36.5 \\
ADORE+STAR 		& 1037 & \textbf{34.4} \\

\hlinew{0.8pt} 
    \end{tabular}
\end{table}

\subsection{Random vs. Hard Negatives}
\label{sec:exp_random_vs_hard}
This section compares random negative sampling and hard negative sampling to answer \textbf{RQ1}.

In Section~\ref{sec:benefit_hard_neg}, we theoretically show that random negative sampling faces a critical problem that some difficult queries may dominate the training process. To verify whether this phenomenon really exists, we plot the distribution of pairwise errors per query in Figure~\ref{fig:dominate} by evaluating the trained Rand Neg model on MARCO Dev Passage dataset. 
The figure shows that $0.2\%$ difficult queries~(pairwise errors $\geq 10^5$) contribute surprisingly $60\%$ of the total pairwise errors.
Therefore, the problem is very serious in practice. 

To investigate whether hard negative sampling alleviates the above problem, we evaluate the total pairwise errors and top-K~(200) pairwise errors for different models and show the results in Table~\ref{tab:optimize_target}. 
As we expect, random negative sampling methods, namely Rand Neg and In-Batch Neg, well minimize the total pairwise errors but cannot effectively minimize the top-K pairwise errors.
The hard negative sampling method, STAR, well minimizes the top-K pairwise errors. The static hard negative sampling methods, namely BM25 Neg and ANCE, achieve compromised top-ranking performance. 
ADORE effectively improves each model's top-ranking performance and performs best using the document encoder trained by STAR. Therefore, the results convincingly show that hard negative sampling can better optimize top-ranking performance. 

\begin{figure}
    \hspace*{\fill}%
    \subcaptionbox{Overlap with the real dynamic hard negatives.
    \label{fig:training_overlap_static}}
    {\includegraphics[width=.47\linewidth]{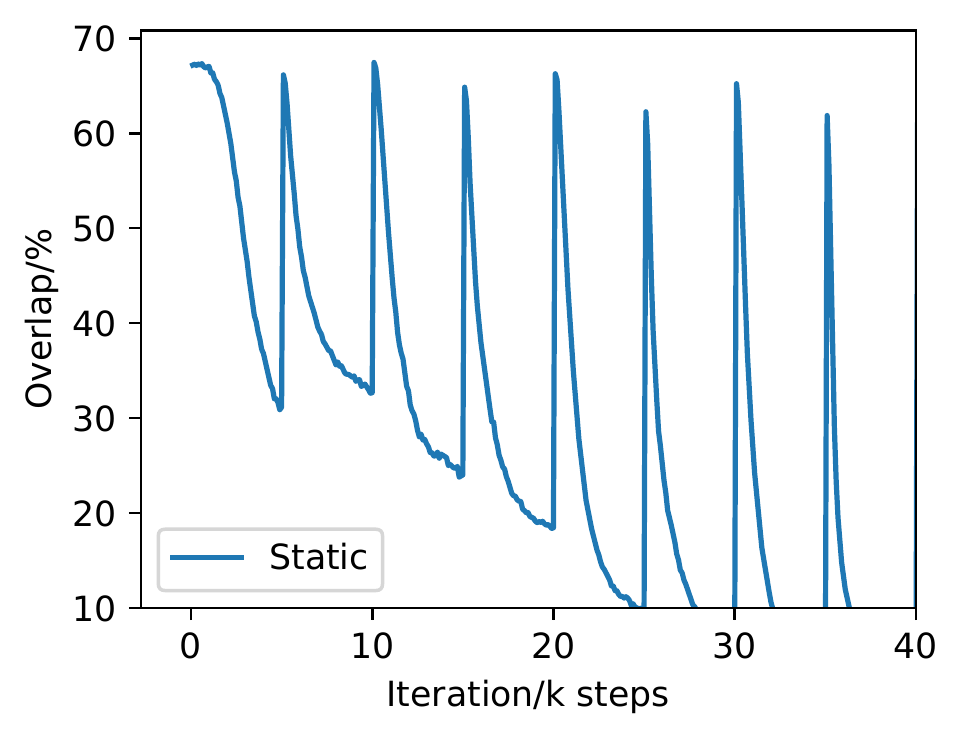}}%
    \hfill\hfill\hfill%
    \subcaptionbox{The MRR@10 at each step. 
    \label{fig:training_mrr_static}}
    {\includegraphics[width=.485\linewidth]{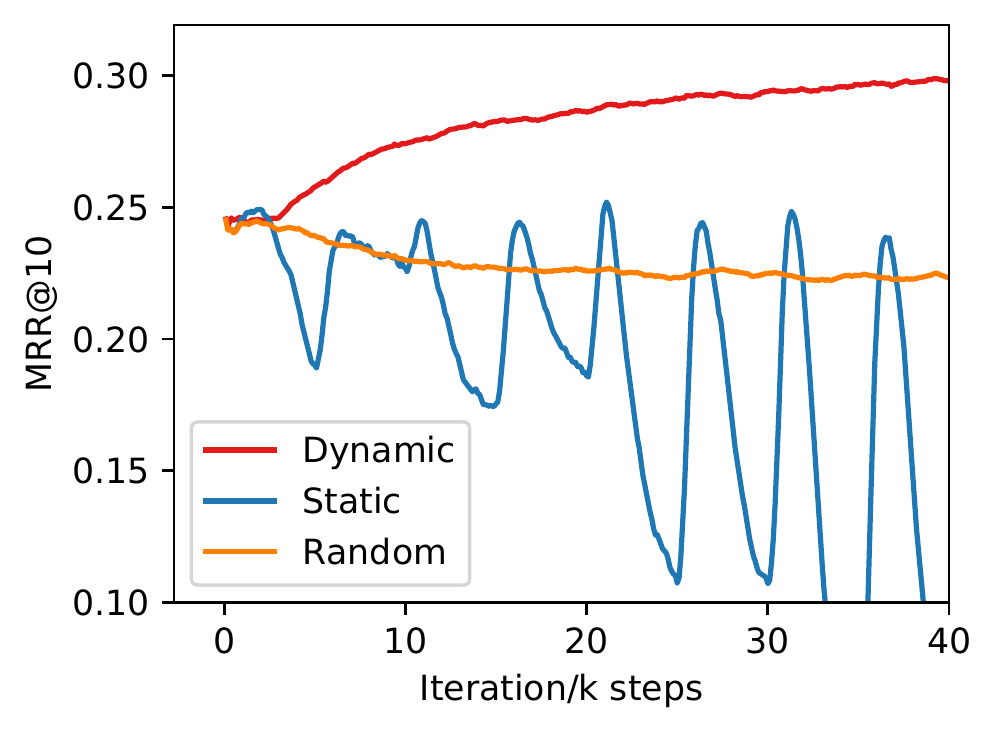}}
    \hspace*{\fill}%
    \caption{The performance of static hard negative sampling on MARCO Train Doc dataset. X-axes is the training steps in thousands. Dynamic/static/random separately denote dynamic hard/static hard/random negative sampling.}
\end{figure}

Note that our experiments contradict previous theoretical analyses. \citet{xiong2020approximate} argued that random and hard negative sampling share the same optimization objective and that the latter's advantage lies in quicker convergence.
However, Table~\ref{tab:optimize_target} shows that random negative sampling well converges and outperforms hard negative sampling in terms of total pairwise errors. 
Figure~\ref{fig:training_mrr_static}, which will be introduced in Section~\ref{sec:exp_static_dynamic}, shows that random negative sampling decreases the top-ranking performance using a relatively good initialization.
Therefore, random and hard negative sampling optimizes different targets, and the benefits of hard negatives are not about convergence speed.

\begin{figure}
    \includegraphics[width=0.6\linewidth, keepaspectratio=True]{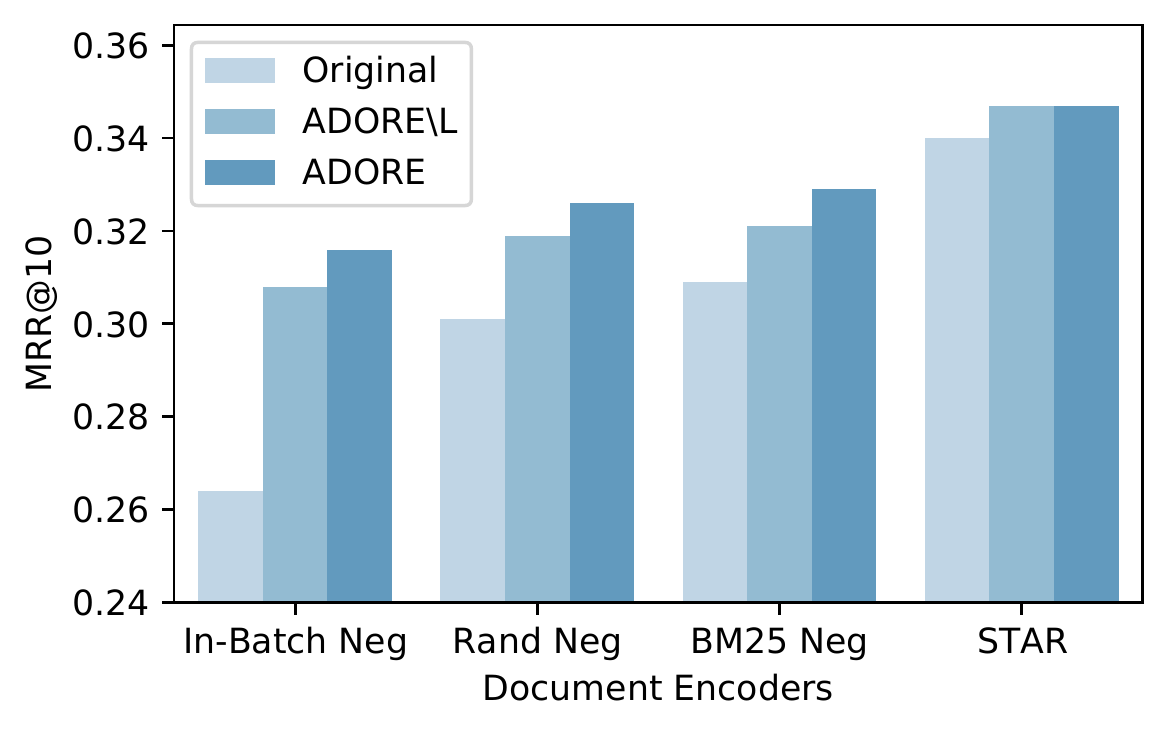}
    \caption{The ablation study for ADORE. ADORE\textbackslash L uses RankNet loss instead of LambdaLoss. Results are MRR@10 values on MARCO Dev Passage dataset.
    } 
    \label{fig:adore_ablation}
\end{figure}

\newcommand\rand{*}
\newcommand\ance{\textsuperscript{$\dagger$}}
\newcommand\adore{\textsuperscript{$\ddagger$}}

\begin{figure}
    \includegraphics[width=0.5\linewidth, keepaspectratio=True]{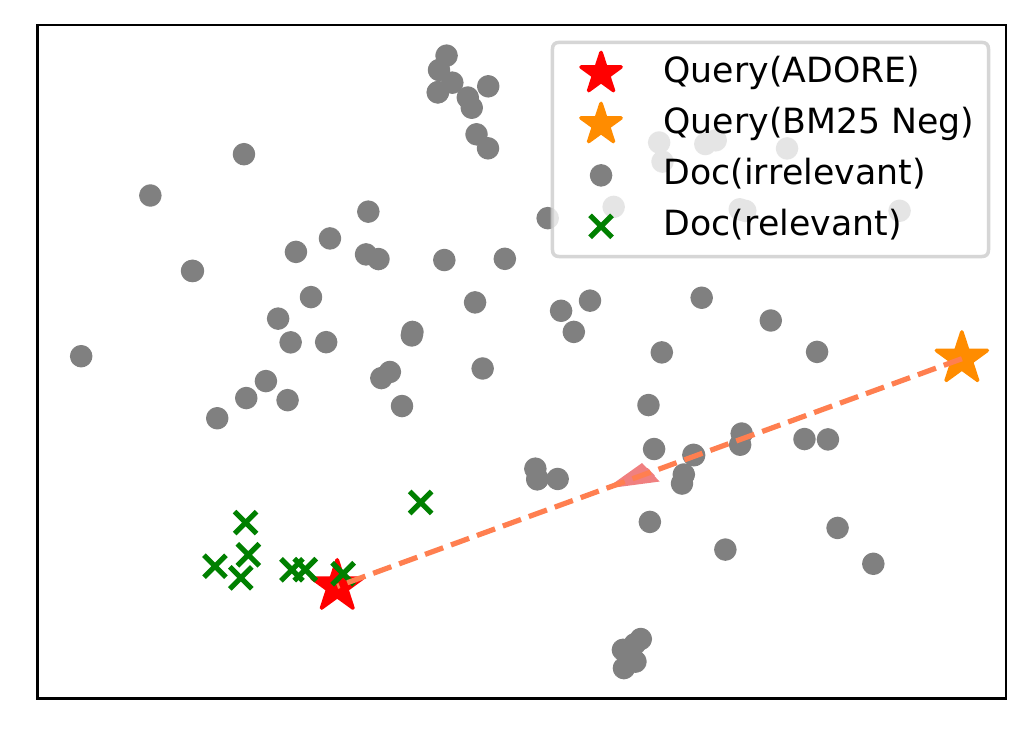}
    \caption{The t-SNE plot of query and document representations for ADORE. The QID is 1129237 and is from TREC DL Doc test set.
    } 
    \label{fig:tsne}
\end{figure}

\begin{table*}[t]
    \centering
    \caption{Results on TREC 2019 Deep Learning Track. 
     We perform the significance test on DR models except for TCT-ColBERT. We use paired t-test with p-value threshold of 0.01 on dev dataset and 0.05 on TREC test dataset. 
     \rand indicate significant improvements over In-Batch Neg, Rand Neg and BM25 Neg.
     \ance indicate significant improvements over ANCE.
     \adore indicate significant improvements over ADORE (In-Batch Neg), ADORE (Rand Neg) and ADORE (BM25 Neg). Best results of DR models are marked bold. Results not available or not applicable are marked as `n/a'. 
     }
    \label{tab:first_stage_effectiveness}
    \begin{tabular}{l|llll|llll} \hlinew{0.8pt}
    & \multicolumn{2}{c}{\textbf{MARCO Dev Passage}} 
    & \multicolumn{2}{c|}{\textbf{TREC DL Passage}} 
    & \multicolumn{2}{c}{\textbf{MARCO Dev Doc}}
    & \multicolumn{2}{c}{\textbf{TREC DL Doc}}
    \\  
    \textbf{Models}
    &  \textbf{MRR@10} &  \textbf{R@100}
    & \textbf{NDCG@10} & \textbf{R@100} 
    &  \textbf{MRR@100} &  \textbf{R@100}
    & \textbf{NDCG@10} & \textbf{R@100} 
    \\ \hline
    
\textbf{Cascade IR}  &  &  &  &  & & & &  \\
{Best TREC Trad LeToR} 
& n/a & n/a & 0.556 & n/a 
& n/a & n/a & 0.561 & n/a \\ 

{BERT Reranker}~\cite{nogueira2019passage} 
& 0.365 & n/a & 0.742 & n/a 
& 0.413 & n/a & 0.646 & n/a 
\\\hline

\textbf{Sparse Retrieval}  &  &  &  &  & & & &  \\

{BM25}~\cite{yang2018anserini}
& 0.187 & 0.670 & 0.506 & 0.453 
& 0.279 & 0.807 & 0.519 & 0.395 \\

{Best TREC Trad Retrieval} 
& n/a  & n/a  & 0.554 & n/a  
& n/a  & n/a  & 0.549 & n/a \\

{DeepCT}~\cite{dai2019context} 
& 0.243 & 0.760 & n/a  & n/a  
& n/a  & n/a  & 0.554 & n/a 
\\\hline

{\textbf{DR: Distillation}}  &  &  &  &  & & & & \\
TCT-ColBERT~\cite{lin2020distilling}
& 0.335 & n/a & 0.670 & n/a 
& n/a & n/a & n/a & n/a 
\\\hline

{\textbf{DR: Negative Sampling}}  &  &  &  &  & & & & \\
In-Batch Neg~\cite{gutmann2010noise} 
& 0.264 & 0.837 & 0.583 & 0.463 
& 0.320 & 0.864 & 0.544 & 0.295 \\

Rand Neg~\cite{huang2020embedding} 
& 0.301 & 0.853 & 0.612 & 0.464 
& 0.330 & 0.859 & 0.572 & 0.284 \\

BM25 Neg~\cite{gao2020complementing} 
& 0.309 & 0.813 & 0.607 & 0.362 
& 0.316 & 0.794 & 0.539 & 0.223 \\ 

{ANCE}~\cite{xiong2020approximate} 
& 0.338\rand\adore & 0.862\rand & 0.654 & 0.445 
& 0.377\rand\adore & 0.894\rand & 0.610 & 0.273 \\ 
\hline 

{\textbf{DR: Ours}}  &  &  &  &  & & & & \\
{STAR} 
& 0.340\rand\adore 	   & 0.867\rand 		  & 0.642 & 0.467\ance
& 0.390\rand\ance\adore & 0.913\rand\ance\adore & 0.605 & 0.313  \\ 

{ADORE+In-Batch Neg} 
& 0.316    & 0.860 		& 0.658\rand & 0.471\ance & 
0.362\rand & 0.884\rand & 0.580 	 & 0.315\ance \\

{ADORE+Rand Neg} 
& 0.326\rand & 0.865\rand & 0.661\rand & 0.472\ance 
& 0.361\rand & 0.885\rand & 0.585 	   & 0.298\ance \\

{ADORE+BM25 Neg} 
& 0.329\rand & 0.846 & 0.661 & 0.431 
& 0.352\rand & 0.872 & 0.610 & 0.293 \\ 

{ADORE+ANCE} 
& 0.341\rand\adore 	   & 0.866\rand 		  & 0.675\rand\ance 		  & 0.454\ance 
& 0.390\rand\ance\adore & 0.902\rand\ance\adore & \textbf{0.634}\rand\ance  & 0.292 \\ 

{ADORE+STAR} 
& \textbf{0.347}\rand\ance\adore & \textbf{0.876}\rand\ance\adore & \textbf{0.683}\rand & \textbf{0.473}\ance 
& \textbf{0.405}\rand\ance\adore & \textbf{0.919}\rand\ance\adore & 0.628\rand 		  & \textbf{0.317}\ance \\ 

\hlinew{0.8pt} 
    \end{tabular}
     
\end{table*}

\let\rand\undefined
\let\ance\undefined
\let\adore\undefined

\subsection{Static vs. Dynamic Hard Negatives}
\label{sec:exp_static_dynamic}
This section compares static hard negatives and dynamic hard negatives to answer \textbf{RQ2}. In Section~\ref{sec:subsec_static_hard_neg}, we argue that the DR model may quickly rank the static hard negatives very low and thus they cannot approximate the dynamic ones. We also analyze that static hard negative sampling does not necessarily optimize ranking performance. This section aims to verify them.

Our experimental settings are as follows. Compared with our theoretical analysis, we further explore the popular periodic index refresh approach by iteratively retrieving static hard negatives every $5k$ training steps~\cite{xiong2020approximate, guu2020realm}. 
The DR model is initialized with a trained BM25 Neg model. 
We fix the document embeddings and retrieve documents for training queries at each step to acquire the dynamic hard negatives and evaluate the ranking performance.
Figure~\ref{fig:training_overlap_static} shows the overlap between static hard negatives and the real dynamic hard negatives.
Figure~\ref{fig:training_mrr_static} presents the ranking performance, which also uses random negative sampling and dynamic hard negative sampling for comparison. 

According to Figure~\ref{fig:training_mrr_static}, the static hard negatives are quickly ranked very low by the DR model and can hardly approximate the dynamic ones as we expect. 
Moreover, the overlap is always less than $70\%$ because the static hard negatives cannot consider the random noise introduced by dropout during training. In this regard, the dynamic hard negatives cannot be entirely replaced by static ones. 

Figure~\ref{fig:training_mrr_static} also supports our analysis by showing the unstableness of static hard negative sampling. The ranking performance fluctuates wildly and periodically. In most time, it significantly underperforms random negative sampling. On the contrary, dynamic hard negative sampling steadily improves the ranking performance.

Though carefully tuning the hyper parameters may alleviate the above problems and several works achieved promising results using this method~\cite{xiong2020approximate, guu2020realm}, the next two sections will show that our proposed methods can better optimize the ranking performance with great efficiency gain.

\subsection{Effectiveness}
\label{sec:exp_effectiveness}
This section investigates the effectiveness of our proposed STAR and ADORE to answer \textbf{RQ3}. 
We conduct experiments on passage retrieval and document retrieval tasks and show the ranking performance in Table~\ref{tab:first_stage_effectiveness}. We discuss the results\footnote{
Although DR models significantly underperform BM25 on TREC DL Doc in terms of R@100, it may be caused by many unlabeled relevant documents~\cite{xiong2020approximate}. 
} in the following.

\subsubsection{Baselines}
Random negative sampling can effectively train DR models compared with sparse retrieval and the LeToR methods.
Rand Neg outperforms BM25, DeepCT, and LeToR even by a large margin on some metrics.  
Static hard negative sampling does not necessarily lead to performance improvements compared with random negative sampling.
It improves the top-ranking performance but may harm the recall capability. 
Specifically, BM25 Neg model almost underperforms Rand Neg in terms of every metric.
ANCE outperforms Rand Neg on MRR@10 but underperforms it in terms of R@100 on the test sets.       

\subsubsection{STAR}
STAR significantly outperforms random negative sampling, especially on top-ranking performance. For example, it outperforms Rand Neg by $13\%$ and $18\%$ on dev passage and dev document sets in terms of MRR@10 and MRR@100, respectively. 
The results are consistent with our original expectation to improve top-ranking performance. 

STAR also significantly outperforms the static hard negative sampling method, especially on recall capability.
For example, it outperforms BM25 Neg and ANCE by $40\%$ and $15\%$ on TREC DL Doc set in terms of R@100, respectively.
Such achievement is very meaningful considering that ANCE iteratively re-builds the index and then updates the static hard negatives while STAR only retrieves once. 
It demonstrates that STAR better optimizes ranking performance through stabilizing the training process. 

STAR outperforms the knowledge distillation approach on the large dev set but underperforms it on the small testing set. 
Therefore, the necessity of knowledge distillation remains further explored since it is usually more computationally expensive than directly training the models.


\subsubsection{ADORE}
\label{sec:exp_effectiveness_adore}
ADORE greatly improves all DR models' performance by further training the query encoders. For example, ADORE improves In-Batch Neg's top-ranking performance by $20\%$ and $22\%$ separately on dev passage and dev doc datasets. It also improves BM25 Neg's recall performance by $19\%$ and $31\%$ separately on testing passage and testing doc sets. 
The results convincingly show the effectiveness of ADORE. 

The combination of ADORE and STAR achieves the best performance. ADORE+STAR greatly outperforms all baselines, especially the existing competitive knowledge distillation approach~(TCT-ColBERT) and the periodic index refreshing approach~(ANCE). 
Furthermore, it nearly matches BM25-BERT two-stage retrieval system on the document retrieval task.

We conduct an ablation study for ADORE to investigate the contribution of dynamic hard negatives and the LambdaLoss. 
The results are shown in Figure~\ref{fig:adore_ablation}.
It demonstrates that using dynamic hard negatives greatly improves the ranking performance, which verifies our previous theoretical analysis. The LambdaLoss can further boost the ranking performance for models like Rand Neg but cannot bring further improvement for STAR. A possible reason is that STAR already emphasizes top-ranking performance compared with methods like Rand Neg.   

To illustrate how ADORE improves ranking performance, we plot a t-SNE example in Figure~\ref{fig:tsne} using a query from TREC DL Doc set. 
ADORE uses the document encoder trained by BM25 Neg and further trains the query encoder. 
After training, ADORE maps the query closer to the relevant documents and thus improves the retrieval performance.

Section~\ref{sec:adore_end_to_end} argues that the optimal DR parameters may be different for different compressed indexes and thus ADORE can achieve better performance through end-to-end training. 
To investigate whether this is true, we use different compressed indexes to train and evaluate DR models. The results are shown in Table~\ref{tab:eval_different_indexes}.
We can see that end-to-end training better optimizes the ranking performance for different compression techniques.  
Thus, ADORE is suitable to improve the performance of compressed indexes.


%

\begin{table}[t]
    \centering
    \caption{
    ADORE+BM25 Neg's ranking performance when using different document indexes for training and evaluation. 
    Results are MRR@10 values on MARCO Dev Passage.
    Row and column names are PQ~\cite{jegou2010product} values, which denote the number of compressed subvectors. 
    A smaller PQ value corresponds to more compression. 
    `-' refers to no compression.}
    \label{tab:eval_different_indexes}
    \begin{tabular}{c|cccc} 
    \hlinew{0.8pt}
    \diagbox[]{\textbf{Train}}{\textbf{Test}}
    & 24 & 48 & 96 & - 
    \\ \hline
{24} & \textbf{0.247} & 0.280 & 0.282 & 0.324 \\  
{48} & 0.244 & \textbf{0.283} & 0.285 & 0.327 \\  
{96} & 0.243 & 0.278 & \textbf{0.291} & 0.326 \\  
{-}  & 0.243 & 0.278 & 0.288 & \textbf{0.329} \\  
\hlinew{0.8pt} 
    \end{tabular}
\end{table}

\begin{table}[t]
    \centering
    \caption{Training efficiency comparison on MARCO Dev passage dataset. All methods use BM25 Neg model as the warm-up model. The training hours and speedup are separately rounded to integers. }
    \label{tab:training_speed}
    \begin{tabular}{l|ccc} \hlinew{0.8pt}
    \textbf{Models}
    & \textbf{GPUs} & \textbf{Hours} & \textbf{Speedup} 
    \\ \hline
    
{ANCE}~\cite{xiong2020approximate} 
& 4 & 645 & \phantom{0}\phantom{0}1x  \\ 

{STAR} 
& 1 & \phantom{0}33 & \phantom{0}20x  \\ 

{ADORE}  
& 4 & \phantom{0}\phantom{0}4 & 179x  \\ 

{ADORE+STAR}  
& 4 & \phantom{0}37 & \phantom{0}18x  \\ 

\hlinew{0.8pt} 
    \end{tabular}
\end{table}

\begin{table}[t]
    \centering
    \caption{ADORE's performance on passage dataset with different compressed indexes. 
     A smaller PQ value corresponds to more compression. 
     Training hours for ${\rm PQ}\!=\!6$ and ${\rm PQ}\!=\!12$ are blank because they are not supported on GPU.
     The last line shows the performance with uncompressed index.
     }
     \label{tab:compressed_index}
    \begin{tabular}{l|cc|ccc} 
    \hlinew{0.8pt}
    & \multicolumn{2}{c|}{\textbf{Index Quality}} 
    & \textbf{Train}
    & \textbf{Dev}
    & \textbf{Test} \\
    
    \textbf{PQ} & \textbf{GB} & \textbf{MRR@10}& \textbf{Hours}  & \textbf{MRR@10} & \textbf{NDCG@10}
    \\ \hline
{\phantom{0}6}
& 0.1 & 0.050 & - & 0.304 & 0.627  \\ 
{12}
& 0.2 & 0.151 & - & 0.318 & 0.635  \\ 
{24}
& 0.2 & 0.221 & 3.0 & 0.324 & 0.644  \\ 
{48}
& 0.5 & 0.254 & 3.2 & 0.327 & 0.652  \\ 
{96}
& 0.9 & 0.273 & 3.7 & 0.326 & 0.656  \\ 
\hline
{--}
& 26 & 0.309 & 3.6 & 0.329 & 0.661  \\ 
\hlinew{0.8pt}
    \end{tabular}
\end{table}

\subsection{Training Efficiency}
\label{sec:exp_train_efficiency}

This section presents the training efficiency of our proposed methods to answer \textbf{RQ3} from two aspects, namely training time and computational resources. Since ANCE is competitive in terms of effectiveness, we use it as our efficiency baseline.

\subsubsection{Training Time} 
\label{sec:exp_train_efficiency_time}
We test the training speed with 11GB GeForce RTX 2080 Ti GPUs and show the results in Table~\ref{tab:training_speed}.
It demonstrates the significant efficiency gains of our proposed methods compared with ANCE. 
The improvement comes from two aspects. 
Firstly, our methods converge very fast. For example, on the passage retrieval task, ANCE needs 600k steps with batch size of 64 while ADORE needs 60k steps with batch size of 32. 
Secondly, to periodically update the static hard negatives, ANCE iteratively encodes the corpus to embeddings and builds temporary document indexes, which takes 10.75 hours each time with three GPUs. In contrast, STAR only builds one temporary index and ADORE does not even have this overhead.
Note that although ADORE retrieves documents at each step, the search is very efficient and takes a total of $40$ minutes, which is about $20\%$ of the entire training time. 

\subsubsection{Computational resources}
\label{sec:exp_train_efficiency_gpu}
This section shows ADORE can greatly save computational resources with compressed document indexes, which is meaningful because GPU memory is usually limited.  
The experimental settings are as follows.
We use the document embeddings generated by BM25 Neg and utilize product quantization (PQ)~\cite{jegou2010product} to compress the index. 
Besides memory usage, we also report the search quality for the compressed index, which is measured with MRR@10 on MARCO Dev Passage dataset. 
The results on shown in Table~\ref{tab:compressed_index}.

We can see that PQ significantly reduces memory consumption and the performance loss is minor.
After compression, ADORE is able to run on only one 11 GB GPU compared with four in Table~\ref{tab:training_speed}.
Specifically, ${\rm PQ}\!=\!96$ reduces the GPU memory footprint to $3\%$ and yields little performance loss (0.326 vs. 0.329).
A smaller ${\rm PQ}$ further reduces GPU memory footprint.
Therefore, ADORE is applicable in a memory-constrained environment.
Note that an over-compressed index (${\rm PQ}\!=\!6$) achieves poor search quality (MRR@10: 0.05), and therefore ADORE almost degenerates into random negative sampling. 
ADORE with ${\rm PQ}\!=\!6$ achieves similar MRR@10 with Rand Neg (0.304 vs. 0.301).

\section{Conclusion}
\label{sec:conclusion}

This paper investigates how to effectively and efficiently train the DR models.
Firstly, we theoretically formalize the training process and compare different training methods. We reveal why hard negative sampling outperforms random negative sampling.
Secondly, we investigate the current popular static hard negative sampling method and demonstrate its risks through theoretical analysis and empirical verification. 
Finally, based on our analysis, we propose two training methods that employ hard negatives to optimize the DR models. 
Experiments on two widely-adopted retrieval datasets show that they achieve significant performance improvements and efficiency gains compared with other effective methods.
Their combination achieves the best retrieval performance. 

There are still some remaining issues for future work. 
Firstly, how to train the document encoder directly based on the retrieval results remains to be explored. 
Secondly, this paper applies DR to ad-hoc search. Future work may examine the proposed methods in other tasks that require a retrieval module, such as Open Question Answering. 

\begin{acks}
This work is supported by the National Key Research and Development Program of China (2018YFC0831700), Natural Science Foundation of China (Grant No. 61732008, 61532011), Beijing Academy of Artificial Intelligence (BAAI) and Tsinghua University Guoqiang Research Institute.
\end{acks}

\bibliographystyle{ACM-Reference-Format}
\balance
\bibliography{references}

\end{document}